\documentclass[12pt]{article}

\usepackage[mathcal]{euscript} 

\usepackage{amsmath}
\usepackage{graphicx,psfrag,epsf}
\usepackage{bm} 
\usepackage{booktabs}
\usepackage{xcolor}
\usepackage{amsfonts}
\usepackage{graphicx}
\usepackage{hyperref}
\usepackage{enumerate}
\usepackage{natbib}
\usepackage{url} 
\usepackage{pifont}
\usepackage{algorithm}
\usepackage{algorithmic}
\newcommand*\CHECK{\ding{51}}
\newcommand{\xmark}{\ding{55}}

\newcommand{\blind}{0}

\graphicspath{{figure/}}

\begin{document}

\def\spacingset#1{\renewcommand{\baselinestretch}%
  {#1}\small\normalsize} \spacingset{1}


\if0\blind { \title{\bf Homogeneity Test for Functional Data based on Data-Depth Plots.} \author{Alejandro Calle-Saldarriaga\thanks{Contact: Alejandro Calle-Saldarriaga; e-mail: \href{mailto:acalles@eafit.edu.co}{acalles@eafit.edu.co}; Affiliation: Department of Mathematical Sciences, School of Science, Universidad EAFIT, Medellín, Colombia, Address: Carrera 49, número 7 sur 50, Medellín, Antioquia, Colombia;  Zip code: 050022} \hspace{.2cm}\\
     Department of Mathematical Sciences, School of Science, Universidad EAFIT\\
    \\
    Henry Laniado \\
    Department of Mathematical Sciences, School of Science, Universidad EAFIT\\
    and \\
    Francisco Zuluaga\\
    Department of Mathematical Sciences, School of Science, Universidad EAFIT}
  \maketitle
} \fi

\if1\blind { \bigskip \bigskip \bigskip
  \begin{center}
    {\LARGE\bf Homogeneity Test for Functional Data based on
      Data-Depth Plots.}
  \end{center}
  \medskip } \fi

\bigskip
\begin{abstract}

  One of the classic concerns in statistics is determining if two
  samples come from the same population, i.e. homogeneity testing. In this paper, we propose a homogeneity test in the context of
  Functional Data Analysis, adopting an idea from multivariate data analysis: the data depth plot (DD-plot). This DD-plot is
  a generalization of the univariate Q-Q plot (quantile-quantile
  plot). We propose some statistics based on these DD-plots, and we
  use bootstrapping techniques to estimate their distributions. We estimate the finite-sample size and power of our test  via simulation, obtaining better
  results than other homogeneity test proposed in the
  literature. Finally, we illustrate the procedure in samples of real heterogeneous data and get consistent results.

\end{abstract}

\noindent%
   \vfill

\newpage
\spacingset{1.5} 

\section{Introduction}
\label{sec:intro}

Functional Data Analysis (FDA) is an area of statistics where
functions are the sample elements. ~\cite{Ramsay1982} coined the
term FDA, but the area is older and dates back to the '50s~\citep{Grenander1950, Rao1958}. With the advance of modern technology, continuously recorded data has become more common and thus interest in the area has spiked. In the FDA context, we
consider that certain functions originated the data that we record
discretely, and that those functions are the sample members, not the
explicit discrete data. Performing some pre-processing steps for smoothing the discrete data is common, but some methods do not need this. For a more complete introduction to
FDA refer to \citep{ramsay2005, ferraty2006}, and refer to
\citep{Wang2016, Cuevas2014} for reviews of the recent advancements in
the area.

FDA usually asks some of the same questions as traditional
statistics. Many times we answer these questions by generalizing the
existing methods of multivariate statistics to the infinite-dimensional context of functions. Some of the questions we have are:
How do we explain this sample using that other sample? How can we
summarize this sample? How do we construct a confidence band for this
statistic? Do these two samples come from the same population? In this paper, we study the last question by proposing a homogeneity test. If two samples come from the same
population they are homogeneous, and they are
heterogeneous if they do not. Lets consider the functional samples
$\mathcal{F} = \{x_1, \dots, x_n\}$ and
$\mathcal{G} = \{y_1, \dots, y_m\}$, defined on the same interval
$\mathcal{T} \subset \mathbb{R}$. We assume that the functions lie on
$C^1(T)$, meaning that their first derivatives are continuous. A
homogeneity test looks to contrast that the two samples come from the
same distribution or not, i.e.
$H_0: \mathcal{F} \stackrel{d}{=} \mathcal{G}$ or
$H_A: \mathcal{F} \stackrel{d}{\neq} \mathcal{G}$, where
$\stackrel{d}{=}$ means equality in distribution. In FDA we rarely
consider explicit distributions, so we say that two samples are
homogeneous when they come from the same `parent' or `generator'
process.

Although many homogeneity tests exist for traditional data (see
\citep{Szkely2002} and \citep{Fong2011} for univariate and
multivariate data respectively), there are few for functional data. On
the best of our knowledge, the homogeneity test for functional data
with the highest power is the one proposed by~\cite{Flores2018}, but
we have evidence that this test has some power problems
in certain specific situations, e.g. when samples only differ
in covariance structure. See section~\ref{sec:sim} for a more detailed
discussion of this.

The main goal of this paper is to construct some homogeneity tests
based on the ideas of multivariate homogeneity using DD-plots proposed
by~\cite{liu1999}, as well as show that this new tests have greater
power than the test proposed by~\cite{Flores2018}, which is the best homogeneity test for
FDA currently in the literature. We evaluate our tests with different
simulated data, where we know a priori if the samples are homogeneous
or not. First, we will use the same simulation scenarios proposed
by~\cite{Flores2018}, but we will also consider other scenarios.

Since distributions of functional data are rarely considered
explicitly, comparing them directly is unfeasible. A good homogeneity
test then must compare different aspects of the sample: for example, a good homogeneity
test may compare means, variance/covariance structure and curve's shape simultaneously, while
a worse homogeneity test may only compare means between samples. We propose a new homogeneity
test that is particularly good at detecting differences between samples in means, variance/covariance
structure or in shape of the curves.

The paper is organized as follows. In section~\ref{sec:depths} we introduce the concept of depth, review different depth measures for functional data and introduce an important homogeneity test found
in the literature. Section~\ref{sec:method}
presents the notion of depth and introduce our depth-based test.  Section~\ref{sec:sim} presents simulation schemes where we compare our tests with others. In section~\ref{sec:data}
we apply the tests to real data. Section~\ref{sec:conc} presents the
main insights of this paper and future research ideas in homogeneity
tests for functional data.

\section{Functional Depths}\label{sec:depths}

In multivariate statistics, depth measures are generalizations of quantiles, as they provide an inward-out ordering of the data. We use similar ideas in FDA, such as centrality, shape, or closeness to other functions to propose depth measures. Different notions of depth for functional data explore different such features. Akin to multivariate depths, we can interpret the deepest function on a sample as the median of that sample. In this section we will consider samples
$\mathcal{F} = \{x_1, \dots, x_n \}$ in an interval $\mathcal{T}$.

~\cite{fraiman2001} introduced the first notion of functional depth. The idea behind this notion is to measure how much ‘time’ each function is deep inside the sample, i.e., how surrounded is the function by other functions. Their idea is to
measure univariate depths for each $t \in \mathcal{T}$, and the
deepest function is the function which on average maximizes this
value.  More formally, fix a $t_o \in \mathcal{T}$ and consider the
univariate depth $D_n(x_i(t_0))$, and define the Fraiman-Muniz depth
as

\begin{equation*}
  FM(x_i(t)) = \int_{\mathcal{T}} D_n(x_i(t)) dt,
\end{equation*}

that is, the average of the depths for all $t_0 \in \mathcal{T}$. Note
that $D_n$ can be any notion of univariate depth. In particular, we
choose

\begin{equation*}
  D_n(x_i(t_0)) = 1 - \left| \frac{1}{2} - F_{n,t}(x_i(t_0)) \right|,
\end{equation*}

where $\hat{F}_{n,t_0}$ is the empirical CDF of
$x_1(t_0), \dots, x_n(t_0)$.

The Random Projection depth \citep{cuevas2007} considers univariate
random projections of the functions. The RP depth is then the average of
the corresponding univariate depths of the projections. More formally,
consider the projection of $x_i$ along direction $v$ as

\begin{equation*}
  r_{i,v}(x_i(t)) = \int_{\mathcal{T}} v(t) x_i(t) dt,
\end{equation*}

and consider $p$ realizations of $v$. The $p$ univariate projections
of $x_i$ are $r_{i,1}, \dots, r_{i,p}$. Consider this notion of
univariate depth for the projections

\begin{equation*}
  d_{i,j} = \min \{ \hat{F}_n(r_{i,j}), 1 - \hat{F}_n(r_{i,j}) \},
\end{equation*}

where $\hat{F}_n$ is the empirical distribution function of all the
functions projected against the same realization of $v$, i.e., if we
are in the $j$th realization of $v$, $\hat{F}_n$ is the empirical
distribution of $r_{1,j}, \dots, r_{p, j}$. Now, the RP
depth is the mean of those univariate depths for each function, i.e.

\begin{equation*}
  RPD(x_i(t)) = \frac{1}{p} \sum_{j=1}^p d_{i, j}.
\end{equation*}

The deepest function will be the one that, on
average, is deeper in its random univariate projections. In this work,
we use $50$ projections of $v$ to compute the RP depth.




~\cite{Nagy2017} warn that traditional functional depths are very simplistic and do not take into account the shapes of the functions or the covariance structure of the sample. They argue that usual functional depths are analogous to taking only coordinate-wise medians in multivariate samples, and interpreting that as the median of the sample, which is ill-advised~\citep{Rousseeuw1999} since the idea of the depths is to
capture global features of the sample's probability distribution
rather than local features~\citep{Paindaveine2013}. Then, they propose
a $J$-th ordered integrated depth that can capture
global features of the sample as

\begin{equation*}
  FD_J(x_i(t); P) = \int_{\mathcal{T}} \dots \int_{\mathcal{T}} D\left((x_i(t_1), \dots, x_i(t_J)^T);P_{\left(X(t_1), \dots, X(t_J)\right)^T}\right).
\end{equation*}

where $P$ is a probability measure in the functional space, $D$ is a
depth measure in a finite space of dimension $J$, and
$P_{ \left(X(t_1), \dots, X(t_J)\right)^T}$ is the corresponding
probability measure in that finite space. To compute $FD_J$, fix a $J$
and plug-in the corresponding empirical measure $P_n$ for
$P_{\left(X(t_1), \dots, X(t_J)\right)^T}$ and select a
$J$-dimensional depth $D$. In this work we set $J=2$ and the
half-space depth proposed by~\cite{Tukey1975}. ~\cite{Nagy2017}
establish connections from the $J$-th ordered depth to the $J-1$-th
derivative of the data: that means that the $2$nd ordered depths
brings us information about the shape of the curve (first derivative)
and the $3$rd ordered depths brings us information about the
convexity of the curve (second derivative).

Other depth functions use different features of the
curves. For example, the h-modal depth generalizes the notion of modes
to functional data \citep{cuevas2006}.~\cite{lopez2009} define the
band depth, which measures how much each curve is contained in the
`band' generated by any other functions in the sample, where a band is
a part of the plane delimited by any number of function. The
deepest function is then the one which is contained in more
bands.

~\cite{Flores2018} define a homogeneity test based on these notions of depths. Their goal is to propose a measure of the distance between samples of functional data.  Let $\mathcal{F}$ and
$\mathcal{G}$ be the two samples we want to test for homogeneity. Denote
$d_{\mathcal{F}}(g)$ the depth of $g$ in a sample
$\mathcal{F} \cup g$, and define
$\mathcal{D}_{\mathcal{F}}(\mathcal{G})$ as the function which
maximizes $d_{\mathcal{F}}(g)$ among $g \in G$. They propose the
following statistics:

\begin{align*}
  P_1(\mathcal{F}, \mathcal{G}) & = d_{\mathcal{F}} \mathcal{D}_{\mathcal{G}} \mathcal{G}, \\
  P_2(\mathcal{F}, \mathcal{G}) & = P_1(\mathcal{F}, \mathcal{F}) - P_1(\mathcal{F}, \mathcal{G}), \\
  P_3(\mathcal{F}, \mathcal{G}) & = d_{\mathcal{F}} \mathcal{D}_{\mathcal{F}} \mathcal{G}, \\
  P_4(\mathcal{F}, \mathcal{G}) & = |P_3(\mathcal{F}, \mathcal{G}) - P_1(\mathcal{F}, \mathcal{F})| |P_3(\mathcal{F}, \mathcal{G}) - P_1(\mathcal{G}, \mathcal{G})|.
\end{align*}

The idea behind $P_1$ is that the function
$\mathcal{D}_{\mathcal{G}} \mathcal{G}$ is the most representative
element of $\mathcal{G}$, so then if its depth is big in
$\mathcal{F}$, it is most likely that $\mathcal{F}$ and
$\mathcal{G}$ are in the same family, i.e., that they are
homogeneous. The idea behind $P_3$ is that if the function of
$\mathcal{G}$ most likely to come from the experiment $\mathcal{F}$ is
very deep in $\mathcal{F}$, then the two experiments will likely be
very mixed and thus come from the same population. $P_2$ and
$P_4$ are normalizations of $P_1$ and $P_3$. Then they use these
statistics and bootstrapping techniques to test the $H_0$ of
homogeneity.

\section{DD-plots and their relation with
  homogeneity}\label{sec:method}

Multivariate data analysis influences FDA. For example, depth measures, first proposed for multivariate data, have been adapted to a functional context, where it has been succesfully applied. For example,
~\cite{Sun2011} proposed a generalization of boxplots for FDA, called
the Functional Boxplot. DD-plots have been also
adapted for FDA but in the context of classification~\citep{Cuesta-Albertos2017}.

~\cite{liu1999} introduced the notion of DD-plots, which are a way of
comparing multivariate distributions of two samples using depth measures. Let $\mathcal{F} = \{x_1, \dots, x_n\}$ and
$\mathcal{G} = \{y_1, \dots, y_m\}$ be the two samples we want to test
for homogeneity, and let $\mathcal{H} = \mathcal{F} \cup \mathcal{G}$.
Define the DD-plot of the combined sample as
$DD(\mathcal{F}, \mathcal{G}, \mathcal{H}) = \{ (D_{\mathcal{F}}(x),
D_{\mathcal{G}}(x)), x \in \mathcal{H} \}$ where $D$ is an arbitrary
depth function in any of the sample spaces $\mathcal{F}$ or
$\mathcal{G}$. Then $DD(\mathcal{F}, \mathcal{G}, \mathcal{H})$ is a
set of pairs of size $|\mathcal{H}| = n + m$. 

\begin{figure}[htbp]
  \centering
  \includegraphics[width=0.5\textwidth]{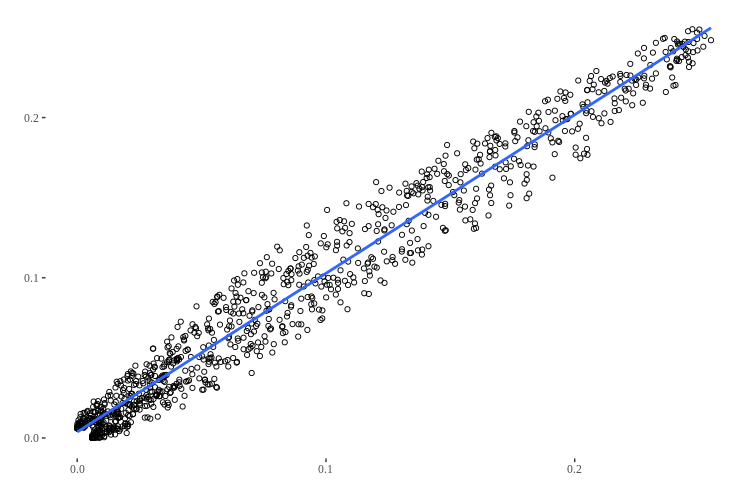}\includegraphics[width=0.5\textwidth]{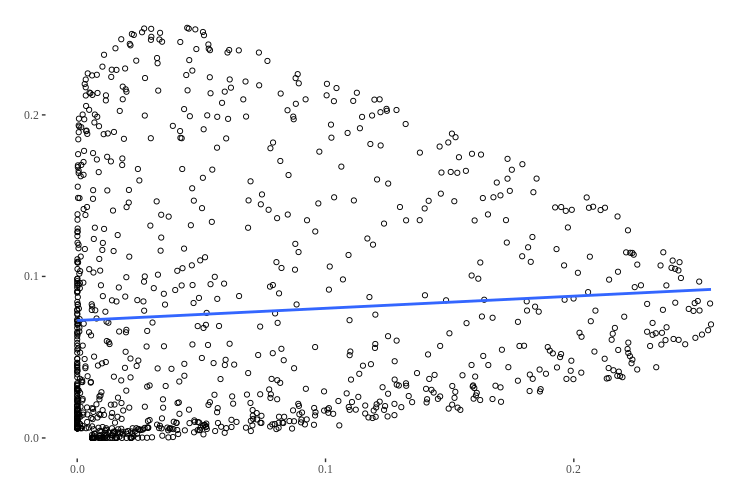}
  \caption[a]{DD-plots for multivariate data. The DD-plot on the left
    is for homogeneous samples, where both samples were generated as
    $N(\bm{0}, I)$. The DD-plot in the right is for heterogeneous
    samples, where the first sample was generated with $N(\bm{0}, I)$ and
    the second sample was generated with $N(\bm{\mu}, \Sigma)$, where
    $\bm{\mu}=[0.5, 1.3]$ and
    $\Sigma = \bigl( \begin{smallmatrix} 1.5 & 0.3 \\ 0.3 &
      1.5\end{smallmatrix} \bigr) $. The Depth function used is the
    Simplicial Depth~\citep{Liu1990}. The homogeneous DD-plot stays concentrated towards the $(0,0)-(1,1)$ line and the heterogeneous is dispersed.}\label{fig:dd-multiv}
\end{figure}

Following this idea, since depths try to
characterize the distributions of samples, they propose that DD-plots
between homogeneous samples must be similar to the identity, that is, the
scatter plot generated by the DD-plot should concentrate towards
the $(0,0)-(1,1)$ line. Figure~\ref{fig:dd-multiv} shows DD-plots for
homogeneous and heterogeneous samples.

Our main idea is to use the proposed relationship presented in ~\cite{liu1999} between homogeneity and DD-plots, and propose some statistics that can capture how concentrated the DD-plot is towards the
$(0,0)-(1,1)$ diagonal. It is worth noting that DD-plots are depth agnostic and you construct them using  any notion of depth. An introduction to the functional depths that we use in this work is in Section~\ref{sec:depths}.

If the same process generated the two samples then the DD-plot should concentrate towards the $(0,0)-(1,1)$ diagonal line. To test this, we consider.

\begin{equation}\label{eq:linear_model}
  D_{\mathcal{F}, i} = \beta_0 + \beta_1 D_{\mathcal{G}, i} + u_i  \quad \forall i = 1, 2, \dots, n + m,
\end{equation}

with $u_i$ being the usual error. We
want to test that $\beta_0 = 0$ and $\beta_1 = 1$, which gives us an idea of how the points concentrate towards the $(0,0)-(1,1)$ diagonal line.

We will be testing using bootstrap-t, which has second-order
convergence properties \citep{Diciccio1996}. Testing with the
traditional t-test is not reasonable as the normality assumption
of $u_i$ does not
hold: the values of $D_{\mathcal{F}, i}$ and $D_{\mathcal{G}, i}$
lie inside $[0,1]$ by definition, so by regressing one against the
other, we will get that $[-1, 1]$ bounds $u_i$, and
thus it is not normally distributed, as the normal distribution
is not bounded. 

Let us say we want to propose a test for a parameter $\theta$, and we 
want to check if $\theta = \theta_0$
Suppose that, with the sample $\bm{X}$, we can estimate $\hat{\theta}$ and
$\hat{\sigma}$, the standard deviation of $\hat{\theta}$. Following
Wald's test, define the test statistic

\begin{equation}\label{eq:T-stat}
  T = \frac{\hat{\theta} - \theta_0}{\hat{\sigma}}.
\end{equation}

We want to estimate a distribution for $T$. Draw bootstrap samples
under the $H_0$ from $\bm{X}$. Call them $\bm{X}^*$, which yields a
$\hat{\theta}^*$ and a $\hat{\sigma}^*$. And with those, we can
estimate the bootstrap replications of $T$ as

\begin{equation}
  T^* = \frac{\hat{\theta}^* - \theta_0}{\hat{\sigma}^*}.
\end{equation}

Now, using this bootstrap-t we will check if $\beta_0 = 0$ and
$\beta_1 = 1$ for the DD-plot. First, compute
$DD(\mathcal{F}, \mathcal{G}, \mathcal{H})$, and get the least
square estimates of the parameters in equation~\ref{eq:linear_model}:
$\hat{\beta}_1 $ and $\hat{\beta}_0$. Also, compute the standard error
of these parameters, given by (See \cite[p.~224]{Seltman2018})

\begin{align*}
  \hat{\sigma}_{\beta_0} & = \sqrt{\frac{\sum_{i=1}^{n+m} \hat{u}_i^2
                           \sum_{i=1}^{n+m} D_{\mathcal{G},i}}{(n+m-2) \sum_{i=1}^{n+m} (
                           D_{\mathcal{G},i} - \bar{D}_{\mathcal{G}})}}, \\
  \hat{\sigma}_{\beta_1}  & = \sqrt{\frac{\sum_{i=1}^{n+m}
                            \hat{u}_i^2}{(n+m-2) \sum_{i=1}^{n+m} (D_{\mathcal{G},i} -
                            \bar{D}_{\mathcal{G}})}},
\end{align*}

where $\hat{u}_i$ are the estimated residuals and
$\bar{D}_{\mathcal{G}}$ is the mean of the $D_{\mathcal{G},i}$. You
can now use equation~\ref{eq:T-stat} to compute

\begin{align*}
  T_0 & = \frac{\hat{\beta}_0}{\hat{\sigma}_{\beta_0}}, \\
  T_1 & = \frac{\hat{\beta}_1-1}{\hat{\sigma}_{\beta_1}},
\end{align*}

which are the test statistics. Then, since in the null hypothesis
$\mathcal{F} \stackrel{d}{=} \mathcal{G}$, resample $\mathcal{H}$ with
replacement to get $\mathcal{H}^*$. Compute the
$DD(\mathcal{F}, \mathcal{G}, \mathcal{H}^*)$ points, and use them to
 compute $T^{*(i)}_0$,
$T^{*(i)}_1$, $i \in \{ 1, \dots, B \}$ bootstrap replicates of each
test statistic.  Now, estimate the p-values of the test as

\begin{align*}
  p_0 & = 2 \min \bigg\{ \frac{1}{B} \sum_{i=1}^B I(T^{* (i)}_0 > T_0),  \frac{1}{B} \sum_{i=1}^B I (T^{* (i)}_0 < T_0) \bigg\}, \\
  p_1 & = 2 \min \bigg\{ \frac{1}{B} \sum_{i=1}^B I(T^{* (i)}_1 > T_1),  \frac{1}{B} \sum_{i=1}^B I (T^{* (i)}_1 < T_1) \bigg\},
\end{align*}

where $I(\cdot)$ is the indicator function.

Select an appropriate confidence level $\alpha$. In this work, we consider $\alpha = 0.05$ Since we have two
p-values, we need to use a sequentially rejective multiple test to
ensure a size $\alpha$ in our test. Using the Holm-Bonferroni method
proposed by~\cite{holm1976}, we order the $p$ ascendingly, getting
$p_{[1]}, p_{[2]}$. Finally, reject the test if $p_{[1]} < \alpha/2$
or $p_{[2]} < \alpha$, and do not reject otherwise. The adjusted
p-value of the test will then be $p = \min \{2 p_{[1]}, p_{[2]} \}$.

\section{Simulation studies}\label{sec:sim}

In this section, we will compute the empirical power and size of the
tests proposed in this work and the test proposed
by~\cite{Flores2018}. For conciseness, from now on we will refer to
that test as `Flores test', using $P_4$ with FM depth, since that was
their best performer, power-wise, in their paper. Let us call as DD-FM test,  DD-RP test, and DD-$FD_J$ the DD-plot based tests with Fraiman-Muniz, Random Projection, and $FD_J$ depths respectively, as described in section~\ref{sec:method}.

We generate samples from two models and count the proportion of rejections. If the same model generated the two samples, this measures finite-sample size, and if two different models generated the two samples, this measures finite-sample power.  The finite-sample size should be near the chosen significance level, and the finite sample-power should be as big as possible.

The models we propose are of the form of equation~\ref{eq:model}, where $\mu(t)$ is the mean function of the
process $x(t)$, $\delta > 0$ is a scalar difference between means of processes; and
$e_i(t)$ is a stochastic Gaussian process with zero mean and
covariance function $\gamma(s,t) = c e^{-c|t-s|}$, where
$s, t \in [0,1]$ and $k, c > 0$. Bigger values of $k$ generate smoother curves, and bigger values of $c$ generate more irregular functions.

\begin{equation}\label{eq:model}
  x_i(t) = \mu(t) + \delta + e_i(t).
\end{equation}

Our simulation procedure is based on the procedure proposed by \cite{Flores2018}. They consider 6 different models, from which they generate samples of 50 curves, and compare samples generated from the first model to samples of the same model and to the other 5. Different parameters in equation 4 generate these different models:

\begin{itemize}
\item Model 0: $\mu(t) = 30t^{3/2}(1-t)$, $\delta = 0$ and
  $\gamma(s,t) = 0.3exp(-3.33|t-s|)$.
\item Model 1: $\mu(t) = 30t^{3/2}(1-t)$, $\delta = 1$ and
  $\gamma(s,t) = 0.3exp(-3.33|t-s|)$.
\item Model 2: $\mu(t) = 30t^{3/2}(1-t)$, $\delta = 0.5$ and
  $\gamma(s,t) = 0.3exp(-3.33|t-s|)$.
\item Model 3: $\mu(t) = 30t{(1-t)}^2$, $\delta = 0$ and
  $\gamma(s,t) = 0.3exp(-3.33|t-s|)$.
\item Model 4: $\mu(t) = 30t{(1-t)}^2$, $\delta = 0$ and
  $\gamma(s,t) = 0.5exp(-5|t-s|)$.
\item Model 5: $\mu(t) = 30t^{3/2}(1-t)$, $\delta = 0$ and
  $\gamma(s,t) = 0.5exp(-5|t-s|)$.
\end{itemize}

\begin{figure}[htbp]
  \centering \includegraphics[width=0.8\textwidth]{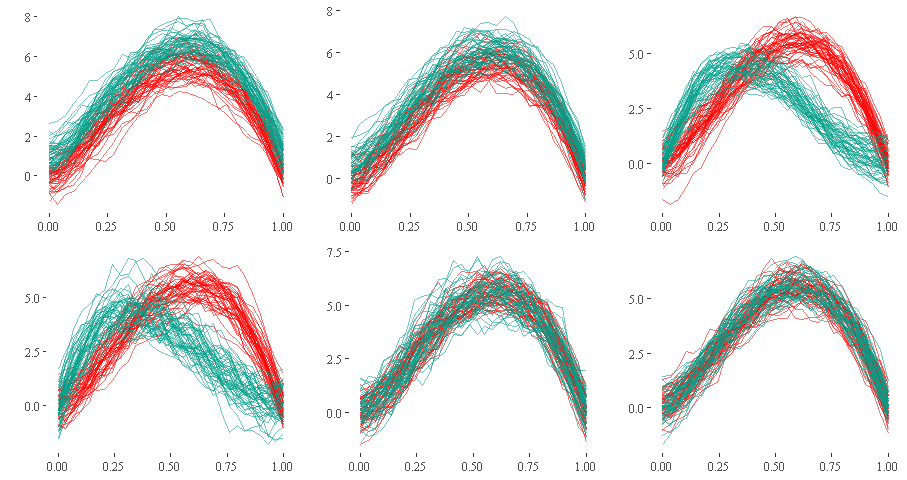}
  \caption{The first five plots are realizations of the reference population in red vs realization of another population in green. The sixth plot is two realizations of model zero. Some changes between samples can be seen easily (first or third plot), but others are quite hard to spot (the last two plots are almost indistinguishable, but the fifth is of heterogeneous samples and the sixth of homogeneous samples).}\label{fig:samples}
\end{figure}

In Figure~\ref{fig:samples} there are samples of the models
described. The first $5$ graphs show two samples generated by different models. The last graph shows two different samples generated from model $0$. A good homogeneity test should reject homogeneity in the first five examples, and not reject homogeneity in the last example.

\begin{figure}[htbp]
  \centering \includegraphics[width=0.8\textwidth]{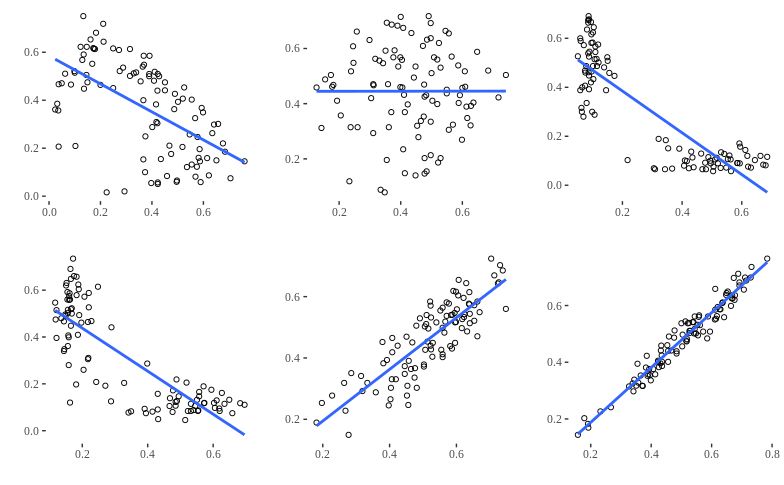}
  \caption{DD-plots using the FM depth for samples of the 6 models
    mentioned, all compared with Model 0.}\label{fig:dd-samples-fm}
\end{figure}

\begin{figure}[htbp]
  \centering \includegraphics[width=0.8\textwidth]{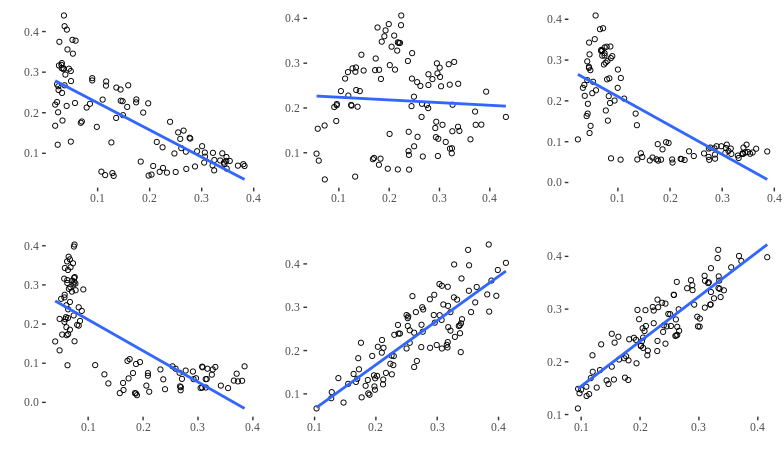}
  \caption{DD-plots using the RP depth for samples of the 6 models mentioned, all compared with Model 0. }\label{fig:dd-samples-rp}
\end{figure}

\begin{figure}[htbp]
  \centering \includegraphics[width=0.8\textwidth]{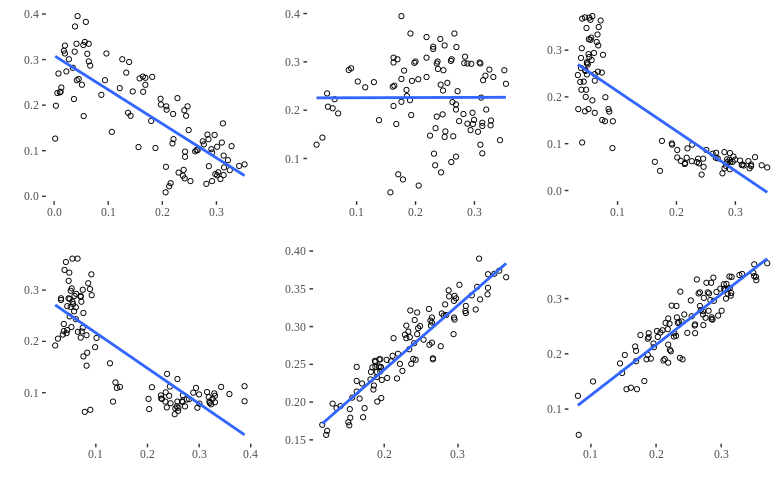}
  \caption{DD-plots using the $FD_J$ depth for samples of the 6 models
    mentioned, all compared with Model 0.}\label{fig:dd-samples-fd1}
\end{figure}

DD-plots for the samples in figure~\ref{fig:samples} can be found in
figures~\ref{fig:dd-samples-fm},~\ref{fig:dd-samples-rp},
and~\ref{fig:dd-samples-fd1}.  In the first four cases, the samples are heterogeneous as the points are not concentrated toward the
the $(0,0)-(1,1)$ diagonal line. It is trickier to say if the samples are heterogeneous in the next two cases: the DD plots seem that of homogeneous samples (and that is not true for the fifth figure), so it is better to perform our proposed homogeneity test to have more certainty.

Table~\ref{tab:summary} summarizes finite sample powers and sizes. All tests achieved good size since all have values near the significance level chose, $\alpha=0.05$. The only test that had an average size bigger than $\alpha$ was the Flores Test, with $0.053$. This means that
DD-plot based tests were better size-wise, with both the DD-FM test and the DD-$FD_J$ test having maximum sizes less than $\alpha$, so it is likely that we will not reject $H_0$ when it is true. Power-wise, the DD-plot based tests were the best performers, with all having a bigger average and minimum power than the Flores’ test. The best performer was the DD-$FD_J$ test: it has the
biggest average and minimum power, meaning that on average it performs very well and that its worst performance was still quite good. Part of its success is that it works very well in instances where the change between samples was in the mean (all tests were able to reject
$H_0$  in these cases), but also could detect when samples differ only in covariance structure, which the other test could not achieve. This could mean that the claims of~\cite{Nagy2017} are true, and this depth can capture global features of the sample’s distribution better than the other depths.

\begin{table}[htbp]
  \centering
  \begin{tabular}{lllll}
    \toprule
    \textbf{Test}                       & \textbf{Average size} & \textbf{Maximum size} & \textbf{Average power} & \textbf{Minimum power} \\ \midrule
    Flores Test   & 0.053        & 0.07         & 0.785         & 0.07 (3v4)\\
    DD-FM         & 0.023        & 0.04         & 0.939         & 0.57 (3v4) \\
    DD-RP         & 0.023        & 0.06         & 0.93          & 0.53 (0v5)\\
    DD-$FD_J$        & 0.023        & 0.05         & 0.968         & 0.82 (3v4) \\\bottomrule
  \end{tabular}
  \caption{Summary of the results for each test. The minimum power
    column also contains which models were being compared when that
    power was obtained. The DD-$FD_J$ has the biggest power and a very good size }\label{tab:summary}
\end{table}

\begin{figure}[htbp]
  \centering \includegraphics[width=\textwidth]{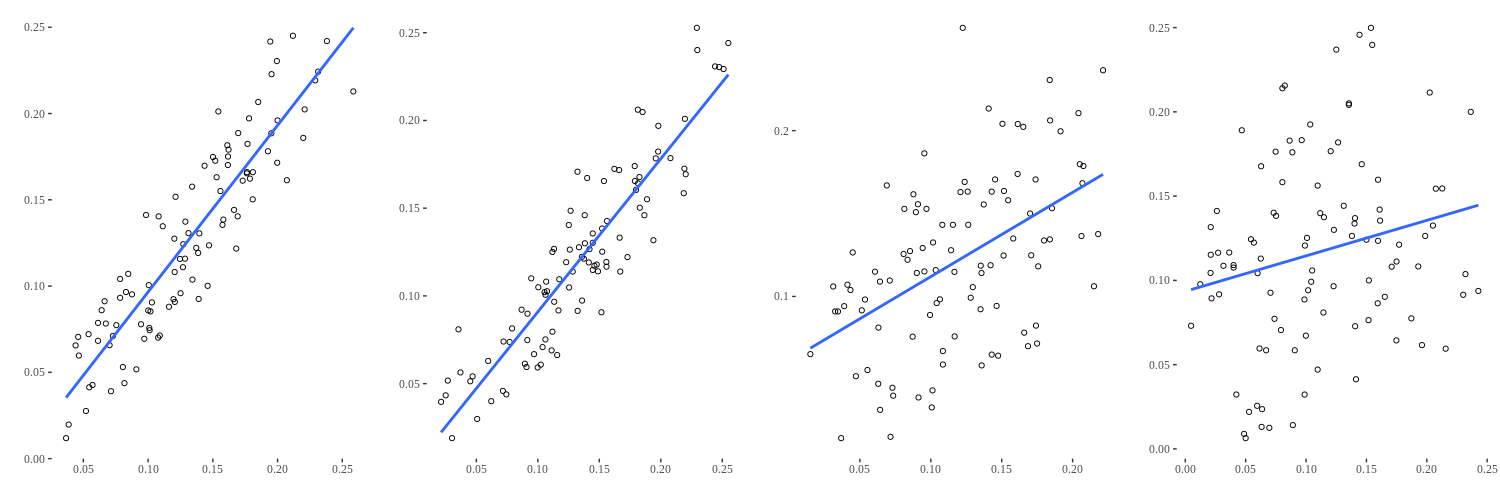}
  \caption{DD-plots with the $FD_J$ depth when $\delta$ is
    increasing.}\label{fig:deltas-dd}
\end{figure}

\begin{figure}[htbp]
  \centering
  \includegraphics[width=0.5\textwidth]{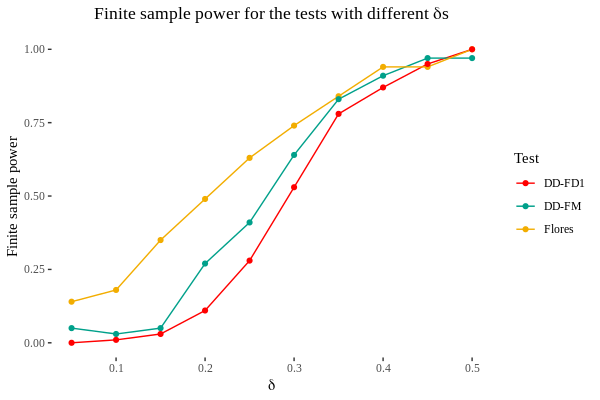}\includegraphics[width=0.5\textwidth]{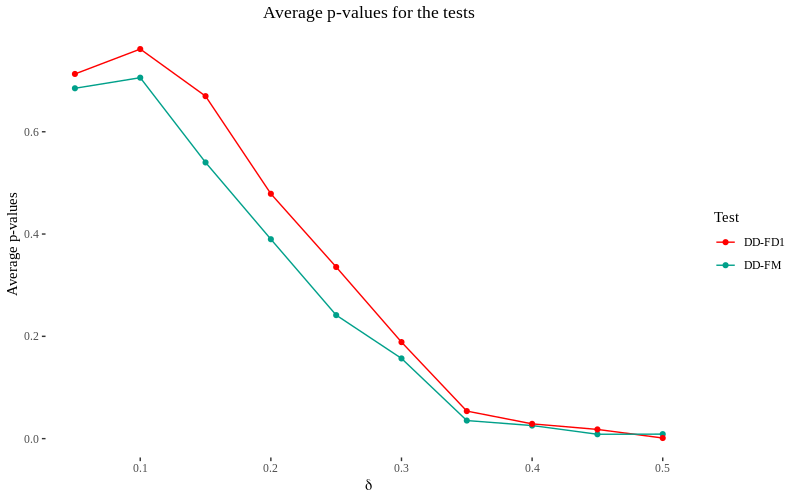}
  \caption{Power of each test when $\delta$ is
    increasing, generating more disperse DD-plots. The Flores Test is able to detect subtle changes in $\delta$ more easily.}\label{fig:pw}
\end{figure}

Some differences between models depend only on certain parameters, e.g. model $0, 1$ and $2$ only differ in their values of
$\delta$. We can then begin changing this parameter to know what is
the breaking point of our tests, i.e., how small can be $\delta$ for
the tests to still detect heterogeneity. DD-plots for different values
of $\delta$ are in Figure~\ref{fig:deltas-dd}, where you can
note that as $\delta$ gets bigger, the DD-plots present greater
deviation from the $(0,0)-(1,1)$ line, which is what you would
expect. Figure~\ref{fig:pw} summarizes the simulation experiments of this scenario. You can see that the Flores Test can
detect earlier the change in $\delta$, meaning that this test is the
best homogeneity test when populations differ in the mean by a certain
scalar. Also, as expected, the average p-values across $100$ tests
tend to be lower as $\delta$ increases.

\begin{figure}[htbp]
  \centering \includegraphics[width=\textwidth]{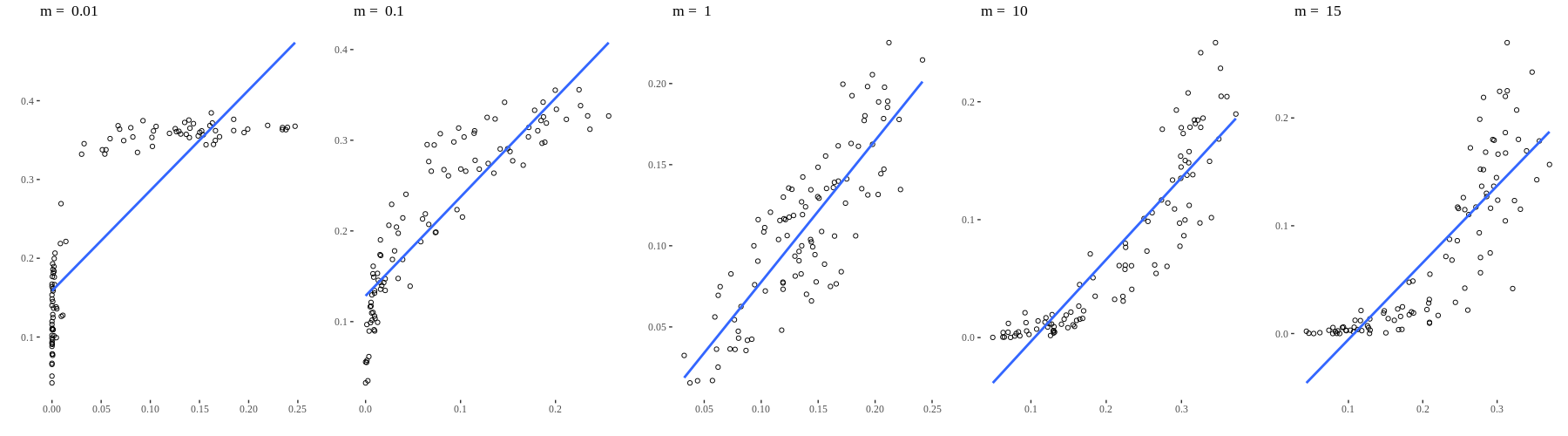}
  \caption{DD-plots with the $FD_J$ depth when $m$ is
    changing. As m decreases or increases much we see some patterns of deviation from the $(0,0)-(1,1)$ diagonal line. }\label{fig:m-dd}
\end{figure}

\begin{figure}[htbp]
  \centering
  \includegraphics[width=0.5\textwidth]{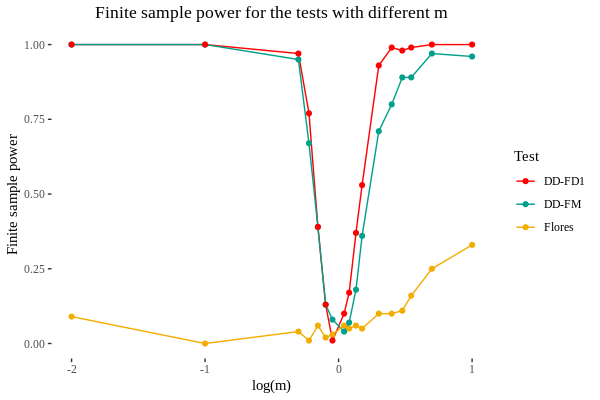}\includegraphics[width=0.5\textwidth]{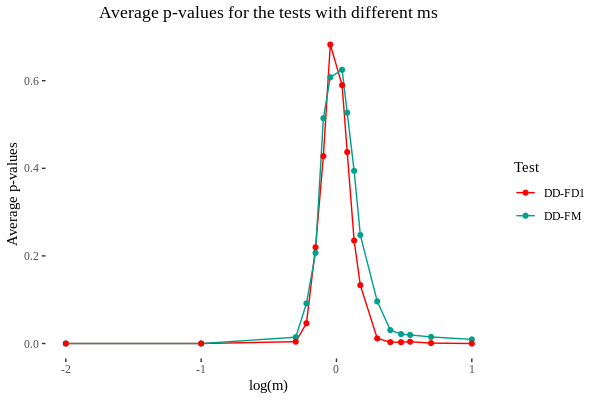}
  \caption{Power of each test when m changes. The Flores Test is not able to detect this changes, and the DD-$FD_J$ is able to detect them quite easily.}\label{fig:pwm}
\end{figure}

On the other hand, models $0, 5$ and models $4, 5$ differ only in the
variance/covariance structure. We can evaluate a similar scenario by changing another parameter: Consider two models that only differ in k, given by equation
\ref{eq:model} , i.e., two models that only differ in covariance structure. The first model will have a fixed $k = 0.3$,
and the second model will have a $k = 0.3m$, with $m$ changing. The
models will be homogeneous when $m=1$, and heterogeneous
otherwise. Let us examine when the different tests have good power with
different values of $m$. In figure~\ref{fig:m-dd} you can see the
change in DD-plots corresponding to the changes in $m$: with small
values or large values of $m$, the plots present behaviors of
non-homogeneous samples, which is what you would expect. Figure ~\ref{fig:pwm} summarizes the simulation experiments of this scenario. You
can see that the Flores Test is not able to detect the
changes in m, while the DD-based tests can get very good
powers with subtle changes in m. The best test in this scenario was
the DD-$FD_J$ test, which reaches perfect sample power quicker than the other tests.

\section{Detecting heterogeneity in samples of real data}\label{sec:data}

In this section, we will consider several data-sets of functional data. They consist of two heterogeneous groups, so our tests should reject the null hypothesis that they are homogeneous. 

\begin{figure}[htbp]
  \centering \includegraphics[width=0.8\textwidth]{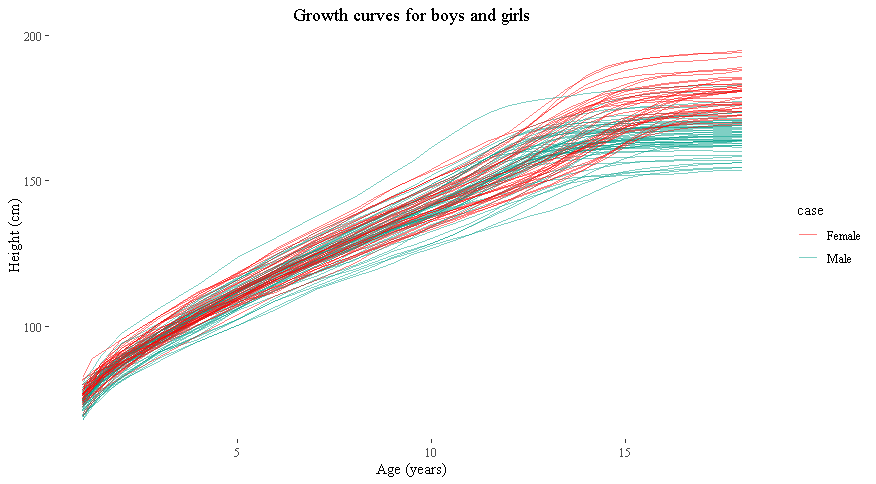}
  \caption{Berkeley Growth Data. Green for males and red for
    females.}\label{fig:heights}
\end{figure}

\begin{figure}[htbp]
  \centering
  \includegraphics[width=0.33\textwidth]{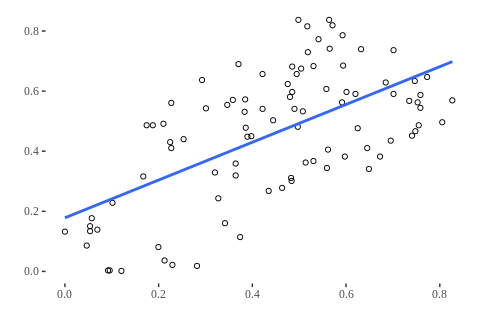}\includegraphics[width=0.33\textwidth]{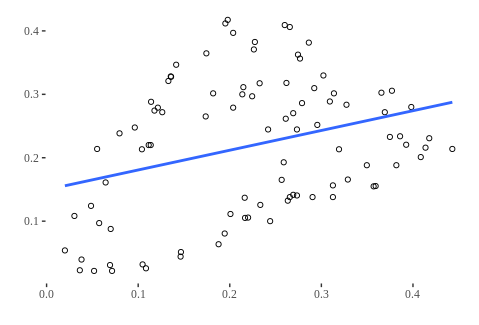}\includegraphics[width=0.33\textwidth]{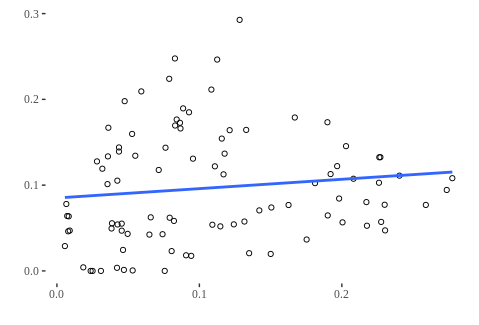}
  \caption{DD plots for the height curves of boys and girls. From left
    to right they use the FM depth, the RP depth and the $FD_J$
    depth.}\label{fig:dd_heights}
\end{figure}

The first set of curves we are considering is the Berkeley Growth
Data, which contains the heights of 39 boys and 54 girls from ages 1 to
18. It is well known that growth dynamics differ from boys to girls,
so our test should reject the null hypothesis of homogeneity. You can
see the curves in figure~\ref{fig:heights}. In
figure~\ref{fig:dd_heights}, you can see the DD-plots obtained from
these two samples using different depths, which show behavior typical
of heterogeneous samples: they are not concentrated towards the
$(0,0)-(1,1)$ diagonal line.

\begin{figure}[htbp]
  \centering \includegraphics[width=0.8\textwidth]{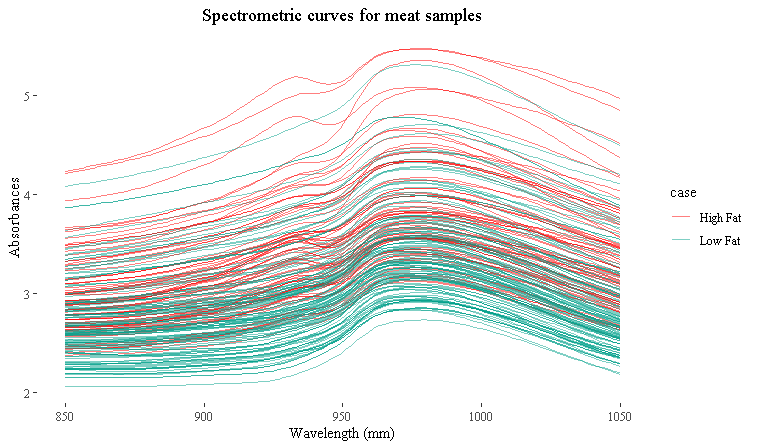}
  \caption{Tecator Spectrometry Data. Green for low-fat meat and red for high-fat meat. }  \label{fig:tecator}
\end{figure}

\begin{figure}[htbp]
  \centering
  \includegraphics[width=0.33\textwidth]{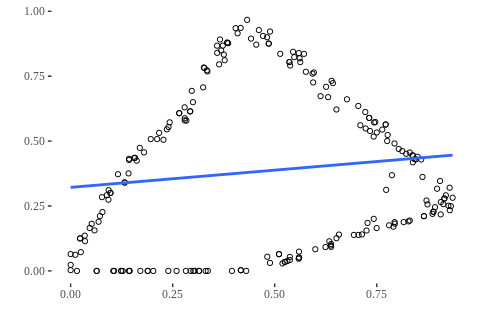}\includegraphics[width=0.33\textwidth]{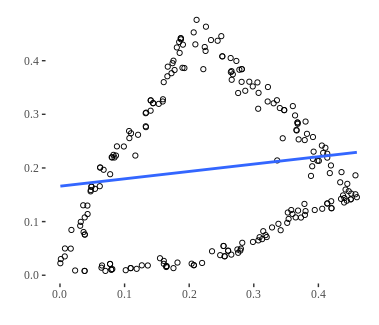}\includegraphics[width=0.33\textwidth]{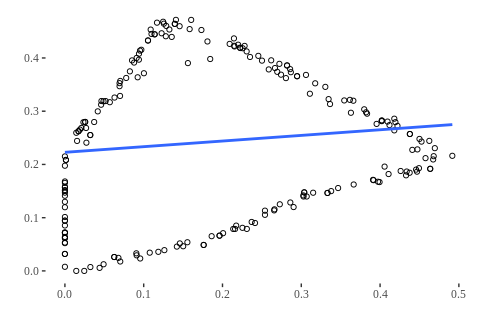}
  \caption{DD plots for spectrometric curves of low and high fat
    containing meats. From left to right they use the FM depth, the RP
    depth, and the $FD_J$ depth.}\label{fig:dd_tecator}
\end{figure}

The Tecator data-set consists of spectrometric curves for chopped pieces of meat, which correspond to the absorbance measured at
$100$ different wavelengths. We can divide these meats into groups, according to~\cite{ferraty2006}: the pieces with small and big fat percentages (where small is less than $20\%$ fat). The curves are in figure~\ref{fig:tecator}. The DD-plots using different depths
of these samples are in figure~\ref{fig:dd_tecator}. The points do not
concentrate on the $(0,0)-(1,1)$ diagonal line.

\begin{figure}[htbp]
  \centering \includegraphics[width=0.8\textwidth]{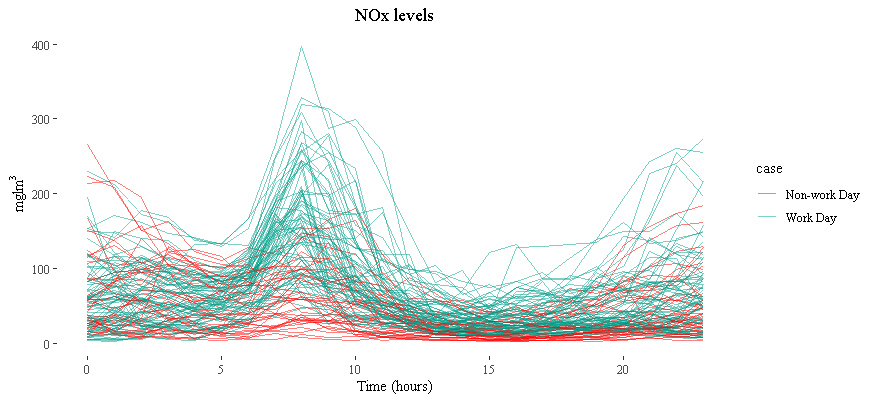}
  \caption{Curves for $NO_x$ levels in a Barcelona Neighborhood for
    work and non-work days.}\label{fig:nox}
\end{figure}

\begin{figure}[htbp]
  \centering
  \includegraphics[width=0.33\textwidth]{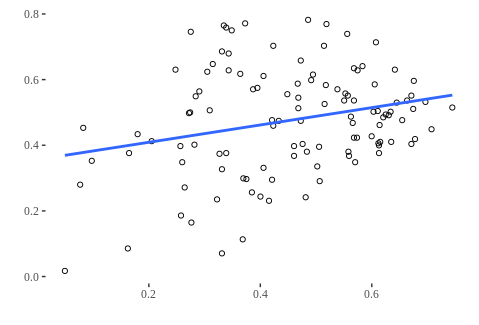}\includegraphics[width=0.33\textwidth]{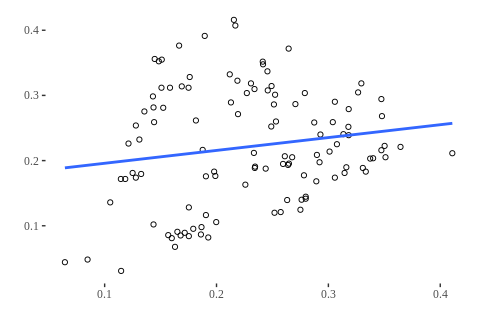}\includegraphics[width=0.33\textwidth]{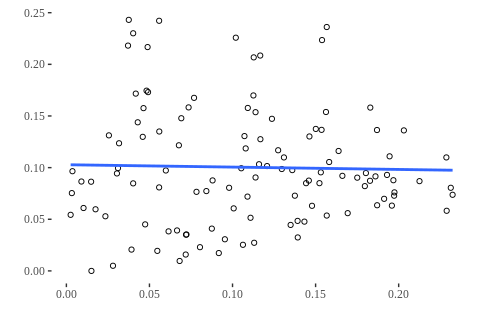}
  \caption{DD-plots for the $NO_x$ curves in Poblenou for work and
    non-work days. From
left to right they use the FM depth, the RP depth, and the FDJ depth.}\label{fig:ddnox}
\end{figure}

~\cite{Febrero2008} propose a dataset measuring nitrogen oxide
emission levels in a control station in Poblenou, a neighborhood in
Barcelona. The station measures $NO_x$ levels in $\mu g/m^3$ every
hour, every day. They split curves into two groups: working days and non-working days. The curves are in
Figure~\ref{fig:nox}, and their respective DD-plots using various depths
are in Figure~\ref{fig:ddnox}. Once more, you can see that the points
are not concentrated toward the $(0,0)-(1,1)$ line.

\begin{figure}[htbp]
  \centering \includegraphics[width=0.8\textwidth]{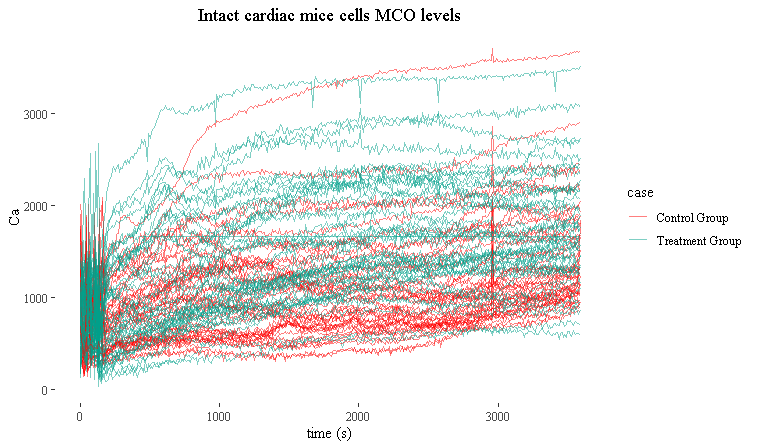}
  \caption{Curves for MCO levels in mouses' cardiac cells. One sample
    is the control sample and the other is treated with a certain
    drug.}\label{fig:mco}
\end{figure}

\begin{figure}[htbp]
  \centering
  \includegraphics[width=0.33\textwidth]{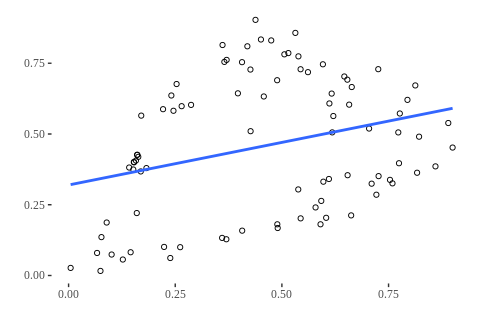}\includegraphics[width=0.33\textwidth]{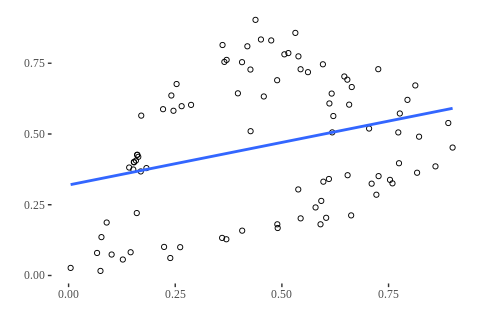}\includegraphics[width=0.33\textwidth]{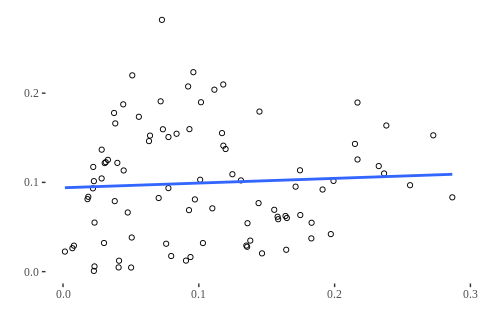}
  \caption{DD plots for the MCO curves in mouse's cardiac cells, one
    group treated with a drug and the other used for control. From
left to right they use the FM depth, the RP depth, and the FDJ depth.}\label{fig:dd_MCO}
\end{figure}

Finally, let us consider the Mitochondrial calcium overload (MCO) data-set. During ischemic myocardia, high levels of MCO relate to better protection against ischemia. Then, it is interesting to see if some drugs can raise MCO levels in rats. This database consists of two groups: one which receives no drug; and another group that receives a drug that can raise MCO levels. Every $10s$ they measure MCO levels. See \citep{ruiz2003}
for the complete description of the experiment. In
figure~\ref{fig:mco} we can see both populations. The DD-plots of
these samples are in figure~\ref{fig:dd_MCO}. The points do not
concentrate on the $(0,0)-(1,1)$ line.

Yet, performing a visual check for homogeneity using DD-plots is not
satisfactory enough, as we do not get important metrics like the
p-value. Performing then the tests proposed in this paper, we get
the results in table~\ref{tab:tests}. The DD-$FD_J$ test was the only
test that was able to reject $H_0$ in all the cases, getting
very small p-values. The Flores Test was only able to reject $H_0$
half the times. Both the DD-FM and the DD-RP tests were able to reject
$H_0$ in all the cases except one.

\begin{table}[htbp]
  \centering
  \begin{tabular}{llllllll}
    \toprule
    \textbf{Data-set} & \textbf{Flores' Test}& \textbf{DD-FM Test} & \textbf{p} & \textbf{DD-RP Test} & \textbf{p} & \textbf{DD-$FD_J$ Test} & \textbf{p} \\
    \midrule
    Heights & \xmark & \xmark & 0.324 & \xmark & 0.268 & \CHECK & 0.00 \\
    Tecator & \CHECK & \CHECK  & 0.034 & \CHECK & 0.096 & \CHECK & 0.00 \\
    MCO & \CHECK & \xmark  & 0.072 & \CHECK & 0.176 & \CHECK  & 0.00 \\
    NOx & \xmark & \CHECK & 0.32 & \CHECK & 0.28 & \CHECK & 0.00 \\
    \bottomrule
  \end{tabular}
  \caption{Result of the tests in the data-sets considered. This \CHECK
    \ means that the test was able to reject $H_0$ and the \xmark \
    means that it did not reject $H_0$. The DD-$FD_J$ was able to detect accurately reject $H_0$ with near $0$ $p$-values always.}\label{tab:tests}
\end{table}

\section{Discussions and conclusions}
\label{sec:conc}

In this article, we proposed a homogeneity test for functional data using DD-plots. We adapted the idea linking multivariate DD-plots and homogeneity, proposed by~\cite{liu1999} to a functional setting. We then formalized these notions into a test that considers a linear model for the DD-plot, and using bootstrap-t, tests two linear hypotheses that relate to homogeneity between samples. We handle the multiple hypotheses using the Holm-Bonferroni method.

This proposal compares in its entirety the depth measures between two samples, making it robust to many scenarios, and this is the advantage of our method. Other tests,
like \citep{Flores2018}, compares only the most representative datum between samples. Also, this test used bootstrap-t procedures, which are second-order accurate \citep{Diciccio1996}, better than the classic bootstrap confidence intervals which are only first-order accurate.

We compared our tests' and other tests' finite-sample performances through simulation.  Our test achieved a desirable finite-sample size, and finite-sample power was greater than the finite-sample power of other tests found in the literature in many different scenarios. The test achieved particularly good results when the difference between samples is only in covariance, and other tests have poor finite-sample power for this scenario.

Our tests results in real data-sets were also satisfactory: in every
case, the results obtained were concordant with reality, while other
tests were not able to do that.

We can extend the methods in this article in several ways: We can change the bootstrap-t for another re-sampling method. Also, we could propose a non-parametric version of the F statistic, used for multiple lineal hypothesis testing. We could implement different depth measures that were not considered here, like~\citep{narisetty2016}, to asses their finite-sample power. We could consider  other simulation scenarios: can our test detect heterogeneity when the samples have the same means and covariance operators but have different kurtosis?~\cite{walter2011} proposes some kurtosis measures for FDA, and we can create new simulation scenarios based on them.

\section{Acknowledgments}

The authors gratefully acknowledge Universidad EAFIT, as this article
is a part of the project \textit{Statistical homogeneity test for
  infinite-dimensional data}, funded by the university.

\bigskip
\begin{center} {\large\bf SUPPLEMENTARY MATERIAL}
\end{center}

\begin{description}

\item[methods.zip:] File containing the implementation of
  \cite{Flores2018} method and of the method described in this
  paper, as well as a script that simulates data as described in this paper. 

\end{description}

\bibliographystyle{chicago}

\bibliography{Bibliography-MM-MC}
\end{document}